\documentclass[12pt,preprint]{emulateapj}
\usepackage{epsfig}
\usepackage{lscape,graphicx,epstopdf}

\shortauthors{Rahmani et al.}

\begin{document}
\def\bv { $b$ ~}
\def\zv { $z$ ~}

\title{THE SPATIAL DISTRIBUTION FUNCTION OF GALAXIES AT HIGH REDSHIFT}

\author{Hadi Rahmani}
\affil{Inter-University Center for Astronomy and Astrophysics, Post
Bag 4, Ganeshkhind, Pune 411007, India; hadi@iucaa.ernet.in}

\author{William C. Saslaw}
\affil{Department of Astronomy, University of Virginia,
    P.O. Box 400325, Charlottesville, VA 22904-4325; Institute of Astronomy, Cambridge, England; wcs@virginia.edu}

\and

\author{Saeed Tavasoli}
\affil{Department of Physics, School of Sciences, Ferdowsi
University, Mashhad, Iran; stavasoli@wali.um.ac.ir}

\begin{abstract}
This is the first exploration of the galaxy distribution function at
redshifts greater than about 0.1.  Redshifts are based on the North
and South GOODS Catalogs.  In each catalog we examine clustering in
the two redshift bands $0.47 \leq z \leq 0.8$ and $0.9 \leq z \leq
1.5$.  The mean redshifts of the samples in these bands are about
$0.6$ and $1.1$. Our main result is that at these redshifts the
galaxy spatial distribution function $f_V(N)$ has the form predicted
by gravitational quasi-equilibrium dynamics for cosmological
many-body systems.  This constrains related processes such as galaxy
merging and the role of dark matter in the range of these redshifts.
\end{abstract}


\section{Introduction}
At relatively low redshifts, $z \lesssim 0.1$, several surveys have
determined the distribution function of galaxies and their
associated dark matter.  These include the Zwicky Catalog
\citep{cra86} the UGC and ESO Catalogs \citep{lah92}, the SSRS
Catalog in three dimensions \citep{fan94}, the Pisces-Persus
Supercluster \citep{sas98} and most recently the 2MASS Catalog with
about 650,000 galaxies \citep{siv05}.   All these catalogs have
yielded the galaxy distribution function, $f_V(N)$, which is the
probability that a volume $V$, or area $A$, placed randomly in space
or on the sky contains $N$ galaxies. This is a very powerful
statistical description of the positions of galaxies both in space
or on the sky. It includes information about the correlation
functions to all orders, as well as characterizing voids and near
neighbor positions, filaments, the average shapes of
clusters, and counts in cells \citep{sas00-b}.

 Under a wide range of conditions, the galaxy distribution evolves
dynamically in quasi-equilibrium and its distribution function has
the form \citep{sas84,sas96,ahm02}

\begin{equation}
f_V(N)=\frac{\bar N (1-b)}{N!} \left [\bar N (1-b)+Nb \right]^{N-1}
e^{-[\bar N (1-b)+Nb]}\label{1}
\end{equation}
This result can be derived from either the thermodynamics or the
statistical mechanics of the cosmological gravitational many-body
system.  The expected number is
\begin{equation}
\bar N = \bar n V\label{2}
\end{equation}
and
\begin{equation}
b = -\frac{W}{2K}\label{3}
\end{equation}
 is the ratio of average gravitational correlation energy to
twice the average kinetic energy of peculiar velocities in the
system.  Equation (\ref{1}) agrees very closely with observational
results in the catalogs mentioned above (with no free parameters),
and also with N-body computer simulations designed to test the
theory on which it is based (reviewed in \citealt{sas00-b}).

The fundamental reason why quasi-equilibrium statistical
mechanics provides a good description of galaxy clustering is that
the cosmological many-body problem is the basic system underlying
this clustering and it contains two different timescales.  One is
the local dynamical timescale on which clustering forms and clusters
interact in overdense regions.  The other is the global timescale
for macroscopic properties to change.  These macroscopic properties
such as density and pressure are averaged over regions which are
large enough to contain a statistically homogeneous distribution of
clusters.  Consequently they initially change on about the Hubble
timescale, but as they gradually virialize over larger and larger
lengthscales, they change even more slowly.  As the N-body
simulations mentioned above have shown, this disparity of timescales
produces quasi-equilibrium evolution of clustering.  (See
\citealt{sas08} for a recent review of this and related topics.)

Of course, there is more to galaxy clustering than the
gravitational interaction of point masses.  Individual galaxy dark
matter haloes have been incorporated into the quasi-equilibrium
theory (\citealt{ahm02}; \citealt{leo04}). Large dark matter haloes
containing many galaxies have been produced in many computer
simulations (eg. \citealt{guo08}). Unfortunately these simulations,
usually depending on many assumptions and parameters, are seldom
compared with equation (\ref{1}) or with the observed spatial and
peculiar velocity distribution functions of galaxies.  So it is
difficult to determine their relevance, even though many detailed
implications of these simulations for galaxies' properties can be
compared with observations.

Although the quasi-equilibrium theory has usually been used
for galaxies of identical masses, it has also  been extended to
systems containing different masses and compared with N-body
simulations \citep{ahm06}. A mass range is a
 secondary effect because most galaxy clustering is produced by the
 mean field of many neighbors rather than by the individual field of
 a nearest neighbor.  The mass-morphology-luminosity relation for
 galaxies may show some dependence of the distribution function on
 mass at low redshifts, but this also appears to be a secondary
 effect \citep{lah92}.  At present, there is not enough morphological
 data for the GOODS or other high redshift catalogs to examine this
 segregation at higher redshifts.

At larger redshifts around $z \simeq 0.5$ and $z \simeq 1.1$, one
might expect to begin to find departures from the form of equation
(\ref{1}), and observations so far have not been used to determine
how well this form of $f_V(N)$ applies at higher redshifts. If it
does apply, then theoretical analyses can predict $b(z)$ in simple
cases \citep {sas86, sas00}.  However several effects could modify
the fundamental nature of equation (\ref{1}). Examples are: 1)
galaxy mergers which could distort $f_V(N)$, 2) the distribution of
dark matter which may have evolved differently from that of luminous
baryonic galaxies, and 3) the initial distribution of galaxies may
have been outside the range (basin of attraction) which could have
evolved into the form of equation (\ref{1}).

Here we report the first observational determinations of $f_V(N)$
for redshifts greater than about $0.1$, in particular for two ranges
$0.47 \leq z \leq 0.80$, and $0.9 \leq z \leq 1.5$.  At these
redshifts, different catalogs based on different magnitude or color
cutoffs will contain different numbers of galaxies.  However,
provided the selection effects are homogeneous over the catalog and
over the sky, they can be normalized out by their value of $\bar N$.
Other more subtle known biases can also be accounted for explicitly
\citep{lah92}. To help minimize these complications it is most
useful to choose samples large enough to provide meaningful
statistics, but over a small enough redshift range to avoid
evolutionary smearing effects. At low redshifts this is easy to do,
but at large redshifts, with presently available data, it involves
practical compromises.

In Section 2 , we describe our samples, and in Section 3 we
determine their distribution functions.  In Section 4 we determine
how the value of $b$ depends on the size of the cells at these
redshifts. Then Section 5 discusses some implications of the
results.
\begin{figure}
\includegraphics[angle=0,clip=,width=2.50in]{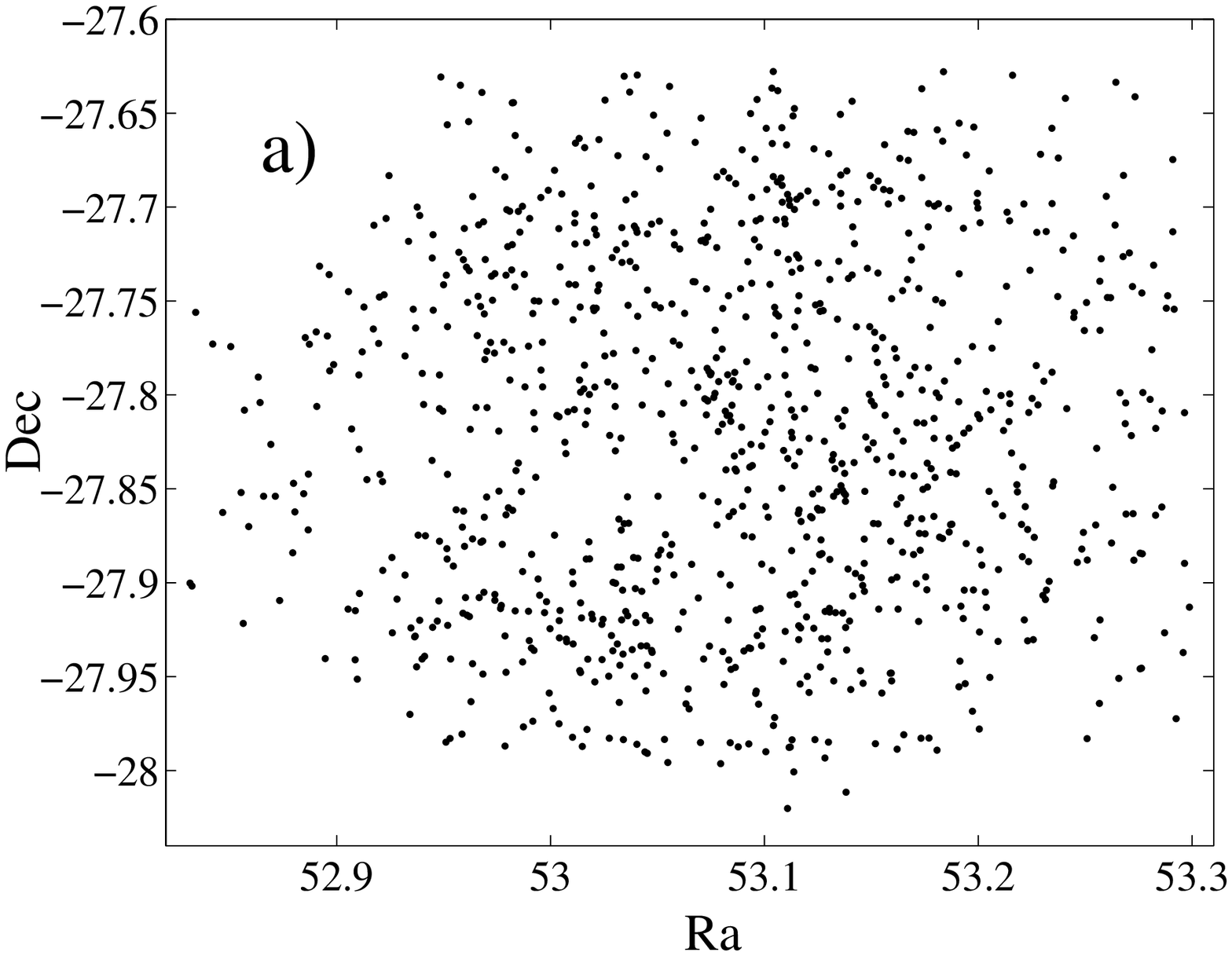}\\
\includegraphics[angle=0,clip=,width=2.50in]{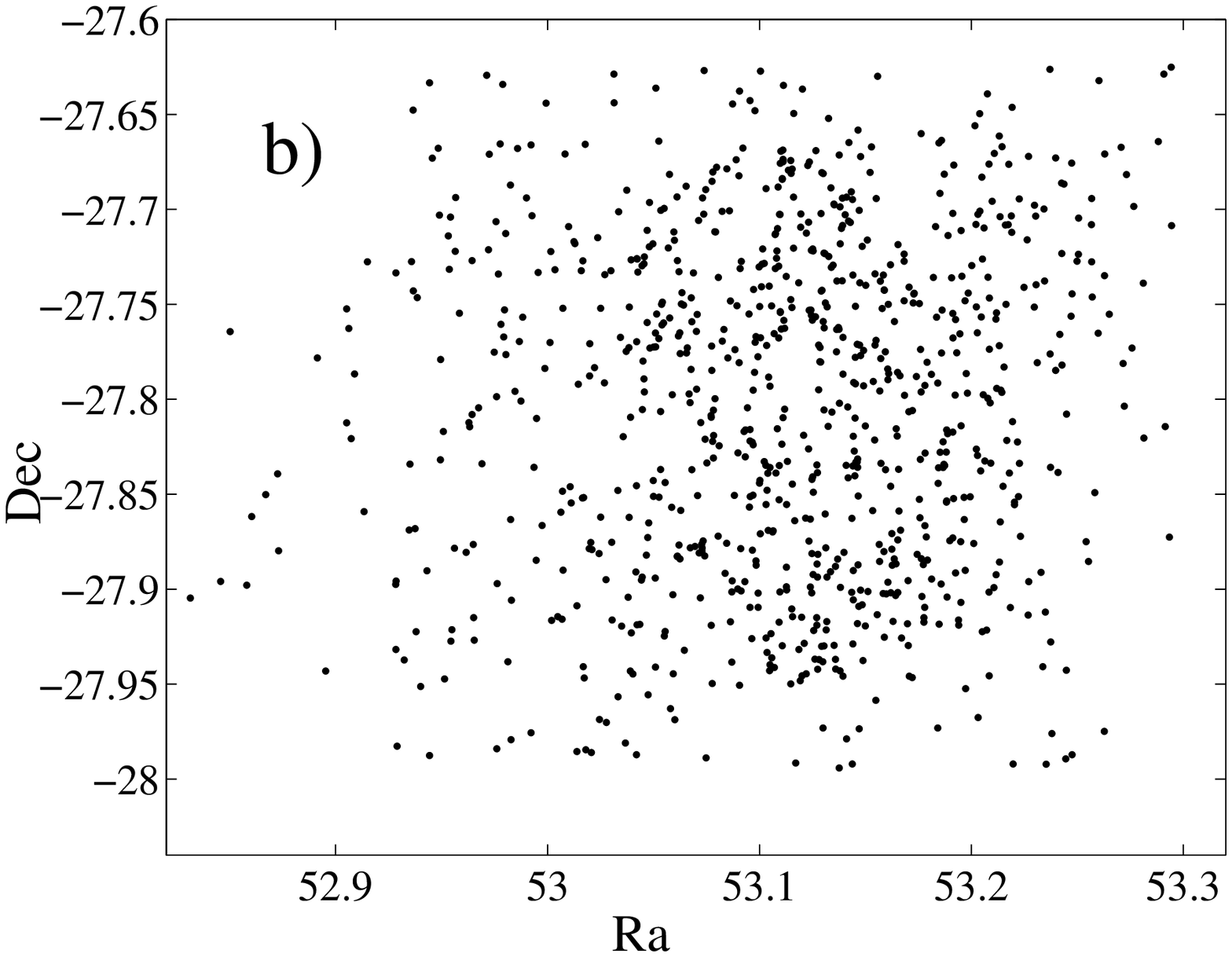}\\
\includegraphics[angle=0,clip=,width=2.50in]{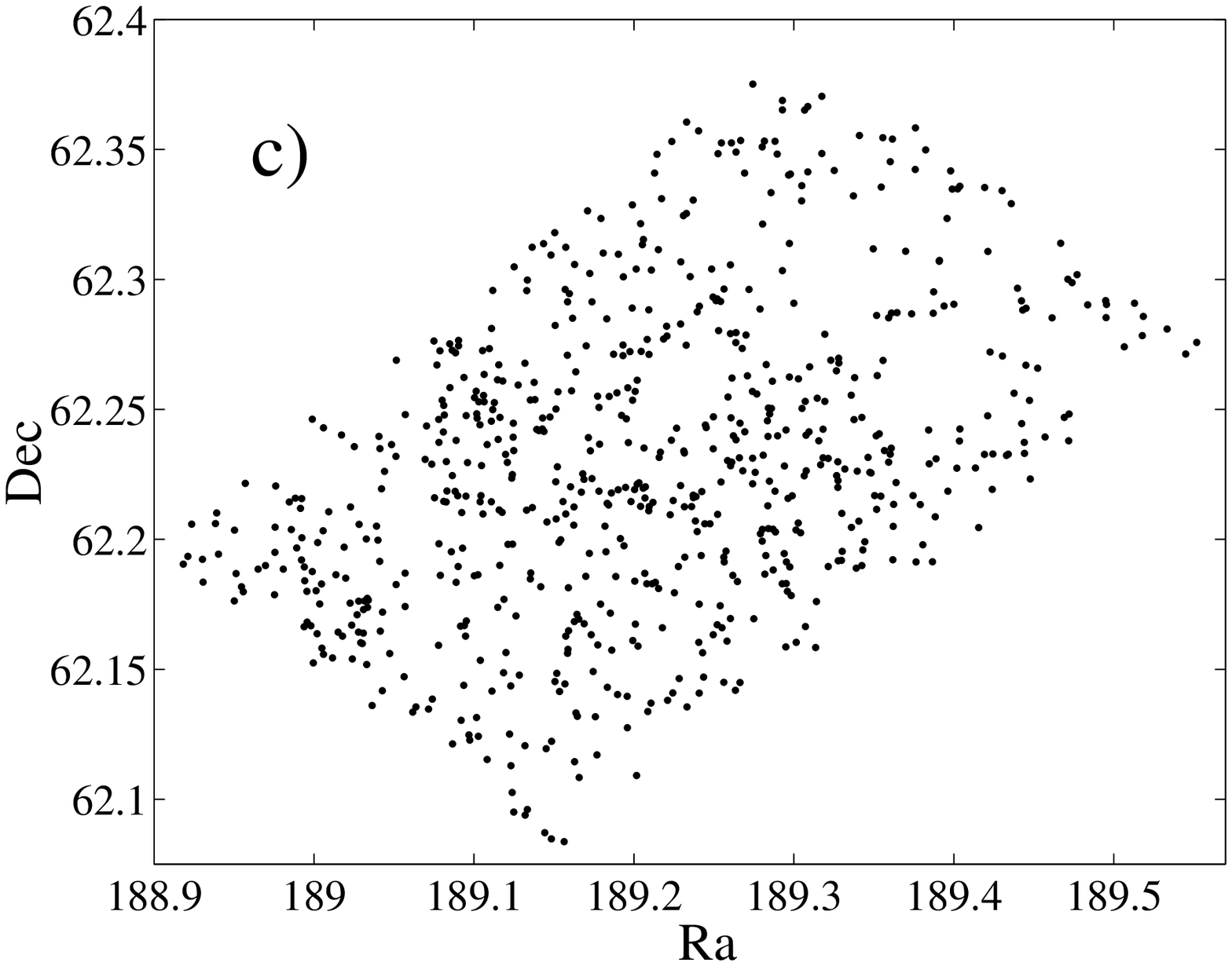}\\
\includegraphics[angle=0,clip=,width=2.50in]{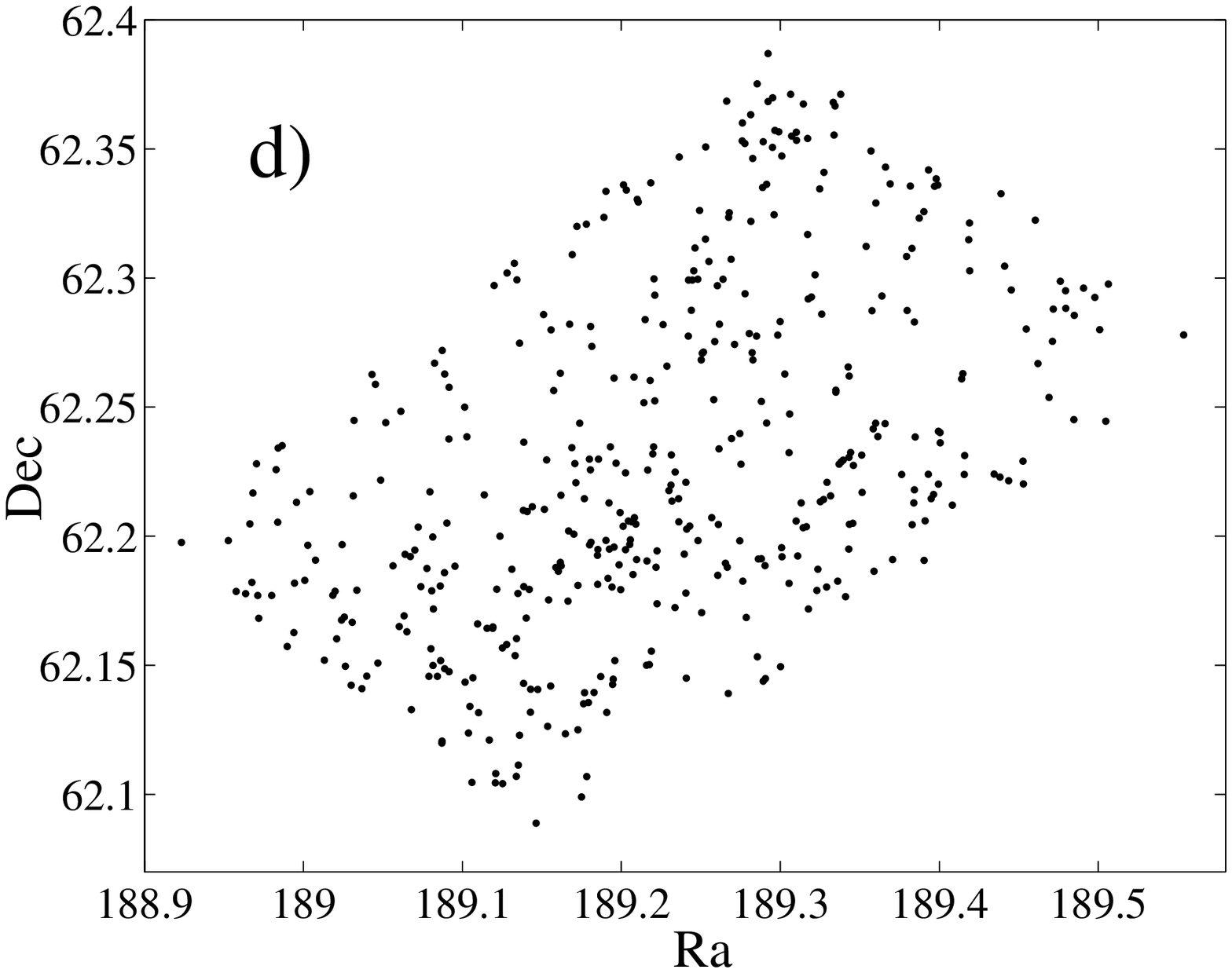}\\
\caption{The distribution of (a) 950 galaxies in the GOODS South
with $0.47\leq z \leq 0.8$, (b) 882 galaxies with $0.9\leq z \leq
1.5$, (c) 685 galaxies in the GOODS North with $0.47\leq z \leq
0.8$, (d) 433 galaxies with $0.9\leq z \leq 1.5$. These are all
projected in equatorial coordinates.\label{fig1}}
\end{figure}

\section{The Observed Samples}
Our analysis is mainly based on the GOODS Catalogs using both its
North and South components separately to check on its homogeneity
and statistical uncertainty.  The GOODS South Catalog has four
components, the VVDS (VIMOS VLT Deep Survey) \citep{lef04}, the ESO1
\citep{van05}, ESO2 \citep{van06}, and a spectroscopic redshift
survey \citep{rav07}, all having spectra taken with VIMOS on the ESO
VLT. These provide 1599, 234, 501, and 961 redshift determinations
for each of the catalogs respectively. We cross correlate these with
a 0.5 arc second search radius to eliminate multiple determinations
of the same redshift. This leaves 950 distinct objects in the
redshift range between 0.47 and 0.8 as well as 882 objects in the
redshift range between 0.9 and 1.5. These ranges were selected to
include a large enough number of galaxies for reasonable statistics
in a small enough redshift band that evolution would be unlikely to
dominate. This is clearly a compromise which could be improved in
future larger but at least equally homogeneous catalogs. Figure
\ref{fig1} (a and b) shows these two redshift samples projected onto
the sky.
\begin{figure}
\includegraphics[angle=0,clip=,width=3.20in]{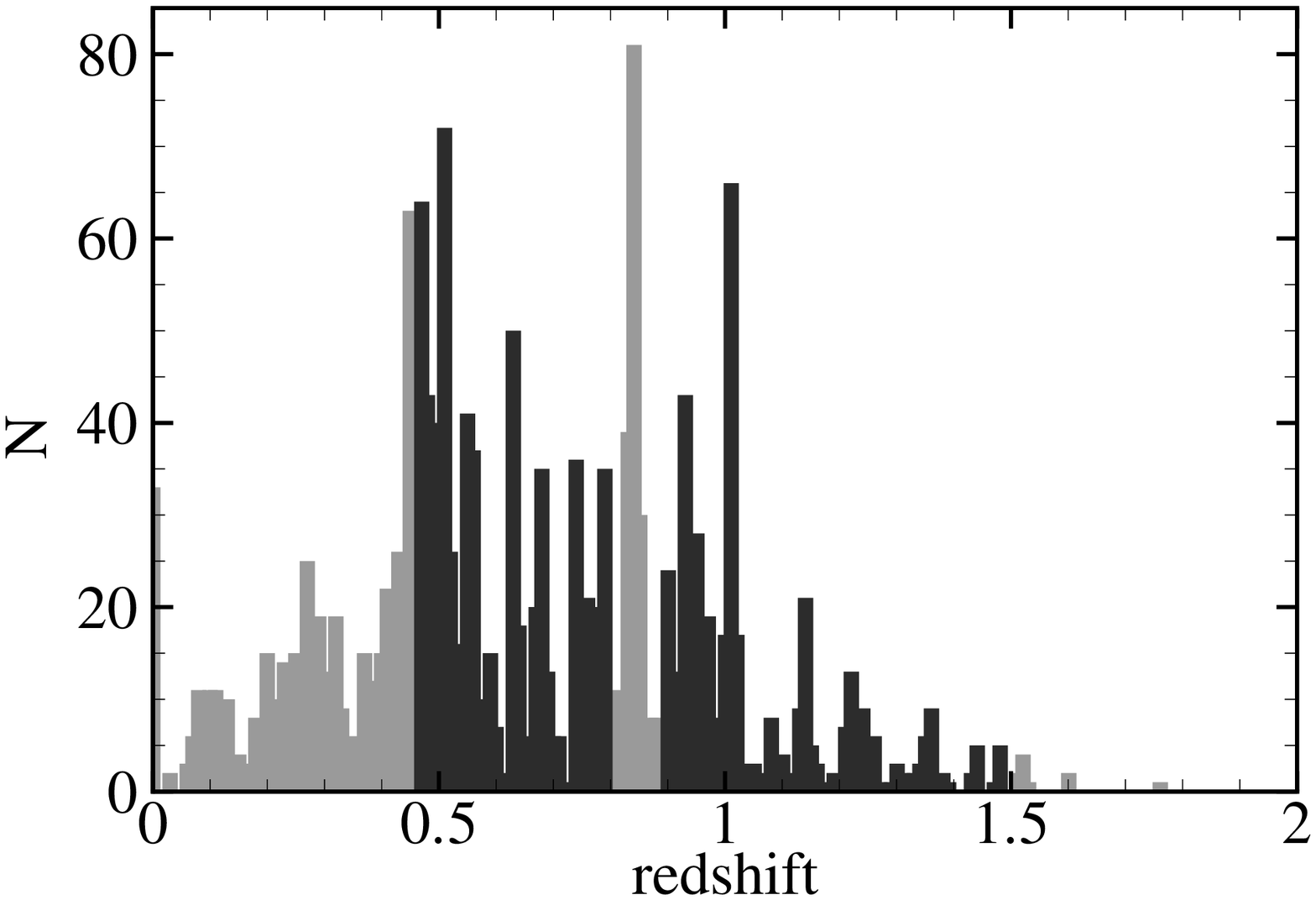}\\
\includegraphics[angle=0,clip=,width=3.20in]{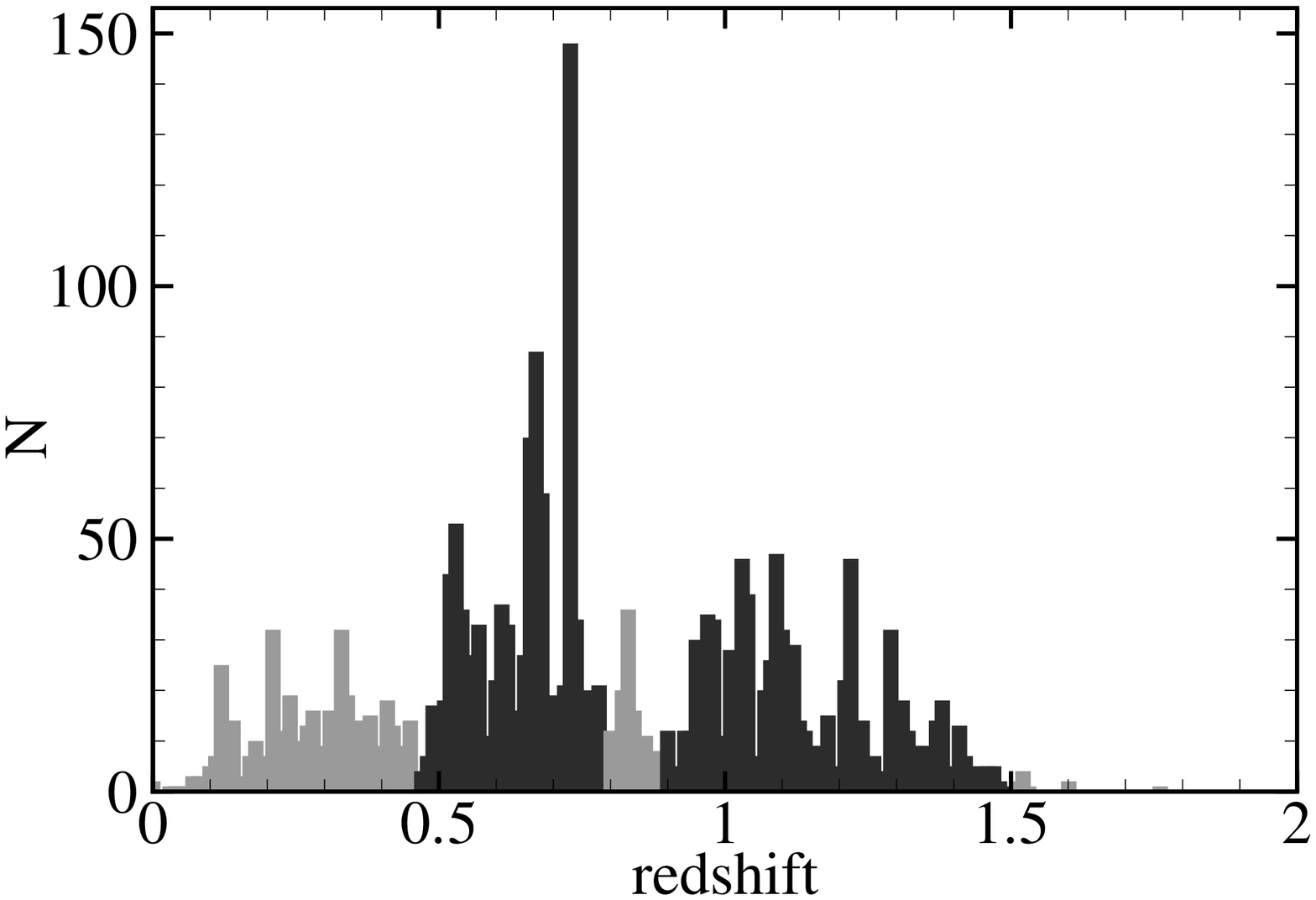}\\
\caption{The redshift distribution of galaxies in North (top) and
South (bottom) catalogs. For each catalog, the galaxies in the
redshift samples we use are shown as the dark filled
histograms.\label{fig3}}
\end{figure}
The GOODS North catalog contains a majority of objects from the Team
Keck Treasury Redshift Survey (TKRS) \citep{wir04}.  They measured
spectroscopic redshifts of 1440 galaxies and AGNs (plus 96 stars) in
this region. There are also 434 other redshifts in this region
obtained by using the LRIS spectrograph \citep {oke95} on the twin
10 m telescopes of the W. M. Keck Observatory \citep
{coh96,coh00,cow96,ste96,low97,phi97,mou97,coh01,daw01} and DEIMOS
spectrograph \citep{fab03} on the Keck twin telescopes \citep
{cow04}. These give a total of 1970 redshifts, of which 685 are
between 0.47 and 0.80 and 433 are between 0.9 and 1.5, comparable
with the numbers in these regions in the South catalog. Figure
\ref{fig1} (c and d) shows their distribution on the sky. Figure
\ref{fig3} shows the redshift distribution for the two ranges in the
North and South, as well as for all redshifts in both regions. The
mean redshifts of the four selected samples are 0.61 and 1.06 in the
North, and 0.65 and 1.14 in the South.

In addition to illustrating the scales of the samples,
figure (\ref{1}) shows that they contain a range of voids,
filamentary structures, underdense regions and clusters.  It also
shows the inhomogeneity of the South samples which contain four
surveys.  In particular, the GOODS South sample for high redshifts
(figure (\ref{fig1}b)) has an average density which increases
systematically with increasing right ascension, even in this small
area.  The GOODS North sample, based mainly on one survey, is more
homogeneous, allowing for greater fluctuations around its smaller
average density.
 
\section{The Galaxy Distribution Functions at $z \simeq 0.63$ and
$z \simeq 1.1$} To obtain the distribution function, we map the
galaxies onto a Hammer-Aitoff equal area projection (e.g.
\citealt{cal02}), divide the area into square cells of a given
angular size and count the number of galaxies in the volume
projected into each cell. The samples are not yet large enough to
examine the areas for completeness and statistical homogeneity in
the usual ways (cf. \citealt{siv05}). However the consistency of our
results below suggests that equation (\ref {1}) will be a  good
representation of the actual distribution function.

\begin{figure*}
\epsscale{1} \plotone{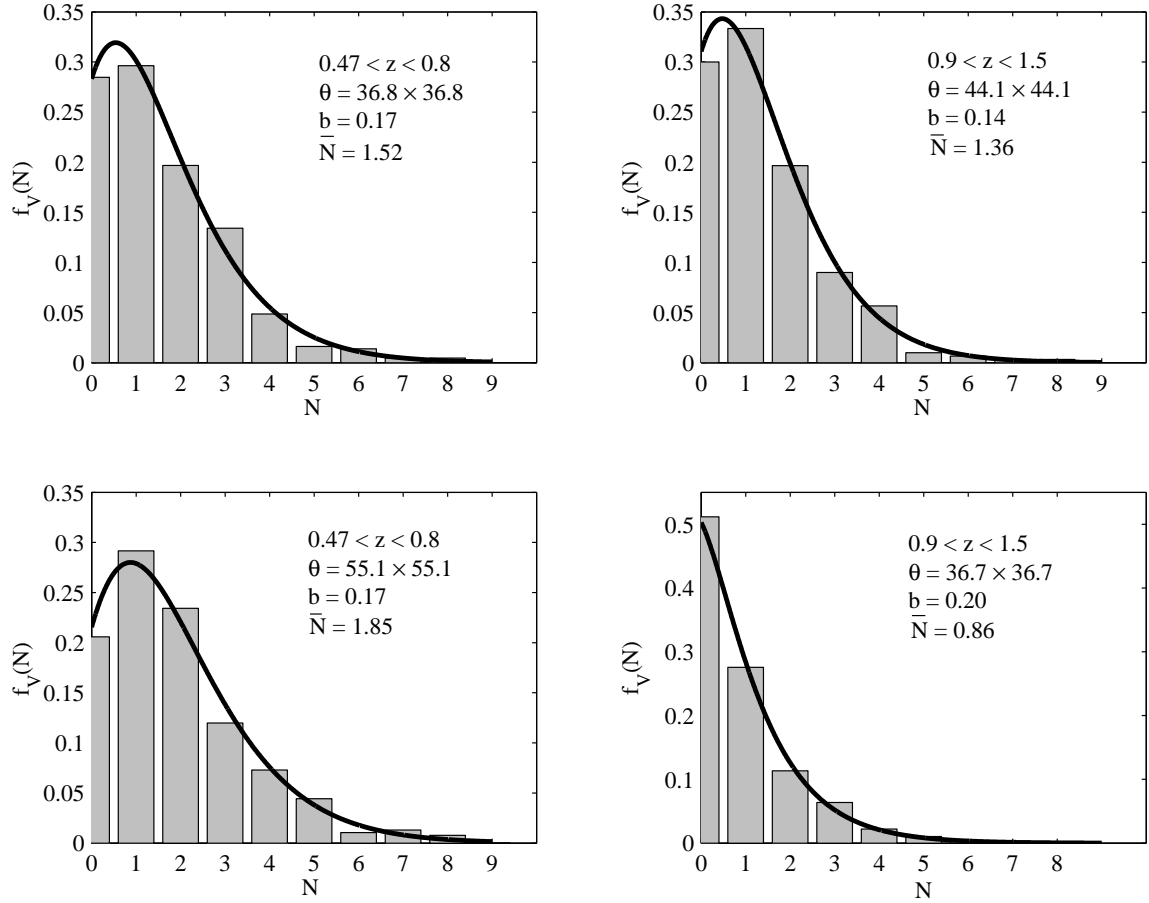} \caption{Theoretical curve (continuous
line) of $f_V(N)$ on the observed histograms for square cells of
GOODS North (top) and GOODS South (bottom) for two redshift
ranges.\label{fig4}}
\end{figure*}

Figure (\ref {fig4}) shows examples of the resulting histograms for
counts of galaxies in cells in the two redshift ranges of both the
North and South Catalogs.  The redshifts and cell sizes are labeled
on the histograms.  Other examples are similar, although as the
cells become larger or the number of cells decreases, the
fluctuations naturally increase.  The solid line is the theoretical
curve of equation (\ref {1}) in which $\bar N$ is determined
directly from the data. The value of $b$ may also be determined
directly from the data using its relation to the variance of counts
in cells having volume V projected onto the sky:
\begin{equation}
\langle(\Delta N)^{2}_{V}\rangle = \frac{\bar
N}{(1-b(V))^{2}}.\label{4}
\end{equation}
Equation (\ref {4}) is the variance of counts in cells and
it follows either directly from equation (\ref {1}) or from its
moments, or from its generating function (see \citealt{sas00-b}).

In addition, we can find $b$ using a least squares fit of equation
(\ref{1}) to the histograms. The values of $b$ for figure (\ref
{fig4}) obtained by least squares fitting are 0.17, 0.12, 0.15, 0.22
from left to right and top to bottom.
 In the Figures we use the values of $b$ from the observed variance of counts
in cells.  Thus the good agreement between observations and theory
is obtained without the use of any free parameters.

Our results in figure (\ref{fig4}), as well as previous
observations of spatial and velocity distribution functions at low
redshifts, are a challenge for the
\begin{figure}
\includegraphics[angle=0,clip=,width=3.20in]{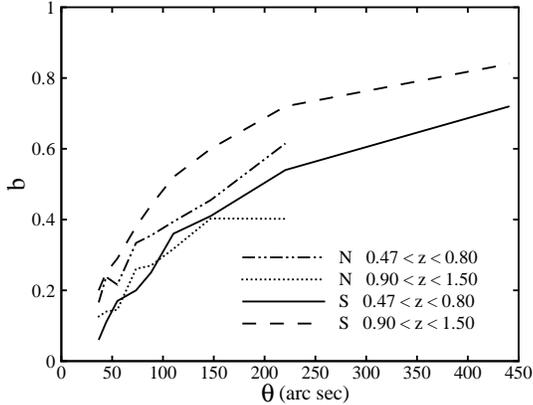} \caption{The value
of $b$ for cells of different angular sizes for the two redshift
samples in the North and South GOODS catalogs.\label{fig5}}
\end{figure}
\begin{deluxetable*}{lccccc|ccccc}
\tablecaption{ Properties of distribution functions from counts in
cells.\label{tbl}} \tablewidth{0pt} \tablehead{
\multicolumn{6}{c}{$0.47 \leq z \leq 0.8$} &
\multicolumn{5}{c}{$0.9 \leq z \leq 1.5$}\\
\hline\\
\colhead{ } & \colhead{$b$} & \colhead{$\bar N$} &
\colhead{$\theta$} & \colhead{$Phy size$} & \colhead{n} &
\colhead{$b$} & \colhead{$\bar N$} & \colhead{$\theta$} &
\colhead{$Phy size$} &
\colhead{n} \\
\colhead{ } & \colhead{ } & \colhead{ } & \colhead{($arc sec$)} &
\colhead{$(kpc)$} & \colhead{(cells)} & \colhead{ } & \colhead{ } &
\colhead{($arc sec$)} & \colhead{$(kpc)$} & \colhead{(cells)} }
\startdata
  &\scriptsize0.17 &  \scriptsize$1.52 $ &  \scriptsize36.8
 &\scriptsize200 &\scriptsize432  &   \scriptsize0.13  &\scriptsize$0.95 $
 &\scriptsize36.8&\scriptsize225 &\scriptsize432\\

&\scriptsize0.24 &   \scriptsize$2.19 $ &  \scriptsize44.1
 &\scriptsize240 &\scriptsize300  &  \scriptsize0.14  &\scriptsize$1.36 $
 &\scriptsize44.1&\scriptsize270&\scriptsize300 \\

 &\scriptsize0.22 &   \scriptsize$3.42 $ &  \scriptsize55.1
 &\scriptsize300 &\scriptsize192  &  \scriptsize0.15  &\scriptsize$2.13 $
 &\scriptsize55.1&\scriptsize337&\scriptsize192 \\

 North&\scriptsize0.33 &   \scriptsize$6.08 $ & \scriptsize73.5
 &\scriptsize400 &\scriptsize108  &  \scriptsize0.26  &\scriptsize$3.79 $
 &\scriptsize73.5&\scriptsize450&\scriptsize108 \\

 &\scriptsize0.36 &   \scriptsize$8.76 $ & \scriptsize88.2
 &\scriptsize480 &\scriptsize75  & \scriptsize0.27  &\scriptsize$5.45 $
 &\scriptsize88.2&\scriptsize540 &\scriptsize75\\

 &\scriptsize0.39 &   \scriptsize$13.69 $ &  \scriptsize 110.2
 &\scriptsize600 &\scriptsize48  &  \scriptsize0.32  &\scriptsize$8.52 $
 &\scriptsize110.2&\scriptsize675 &\scriptsize48\\

 &\scriptsize0.46 &   \scriptsize$24.3 $ & \scriptsize147.0
 &\scriptsize801  &\scriptsize27  &  \scriptsize0.40  &\scriptsize$15.2 $
 &\scriptsize147.0&\scriptsize900 &\scriptsize27\\

 &\scriptsize0.62 &   \scriptsize$54.75 $ &  \scriptsize220.5
 &\scriptsize1201  &\scriptsize12 & \scriptsize0.40  &\scriptsize$34.1 $
 &\scriptsize220.5&\scriptsize1349&\scriptsize12 \\
 &  &   &  & &
  &
&    & & &\\
 \hline &  &   &  & &
  &
&    & & &\\

 &\scriptsize0.06 &  \scriptsize$0.82 $
&  \scriptsize36.7
 &\scriptsize205 &\scriptsize864  &  \scriptsize0.20  &\scriptsize$0.86 $
 &\scriptsize36.7&\scriptsize226&\scriptsize864 \\

 &\scriptsize0.11 &   \scriptsize$1.18 $ &  \scriptsize44.1
 &\scriptsize246  &\scriptsize600 &  \scriptsize0.25  &\scriptsize$1.20 $
 &\scriptsize44.1&\scriptsize271&\scriptsize600 \\

 &\scriptsize0.17 &   \scriptsize$1.85 $ &  \scriptsize55.1
 &\scriptsize308  &\scriptsize384 &  \scriptsize0.29  &\scriptsize$1.90 $
 &\scriptsize55.1&\scriptsize339 &\scriptsize384\\

 &\scriptsize0.20 &   \scriptsize$3.30 $ &  \scriptsize73.5
 &\scriptsize410  &\scriptsize216 &  \scriptsize0.38  &\scriptsize$3.40 $
 &\scriptsize73.5&\scriptsize452 &\scriptsize216\\

 South&\scriptsize0.25 &   \scriptsize$4.70 $ &  \scriptsize88.2
 &\scriptsize492 &\scriptsize150  &  \scriptsize0.44  &\scriptsize$4.90 $
 &\scriptsize88.2&\scriptsize542 &\scriptsize150\\

 &\scriptsize0.36 &   \scriptsize$7.40 $ &  \scriptsize110.2
 &\scriptsize615 &\scriptsize96  &  \scriptsize0.52  &\scriptsize$7.70 7$
 &\scriptsize110.2&\scriptsize677&\scriptsize96 \\

 &\scriptsize0.41 &   \scriptsize$13.10 $ &  \scriptsize147.0
 &\scriptsize820  &\scriptsize54 &  \scriptsize0.60  &\scriptsize$13.70 $
 &\scriptsize147.0&\scriptsize903 &\scriptsize54\\

 &\scriptsize0.54 &  \scriptsize$29.60 $ & \scriptsize220.5
 &\scriptsize1230 &\scriptsize24  &  \scriptsize0.72  &\scriptsize$30.80 $
 &\scriptsize220.5&\scriptsize1355&\scriptsize24 \\

 &\scriptsize0.72 &  \scriptsize$118.30 $ &  \scriptsize441.0
 &\scriptsize2460 &\scriptsize6  & \scriptsize0.84  &\scriptsize$123.20 $
 &\scriptsize441.0&\scriptsize2710&\scriptsize6
\enddata
\end{deluxetable*}
usual computer simulations to
reproduce (without too many free parameters).  For example, one of
the important consequences of figure (\ref{fig4}) is the role of
galaxy mergers in altering their distribution function over time.
From the viewpoint of the theory behind equation (\ref{1}), the
robustness of $f_V(N)$ to mergers can be understood analytically
\citep{yan08}.
\section{The Observed Dependence of \bv on Scale and Redshift}
The value of $b$ is known to depend on the size of the cells for
which it is measured.  If cells are so small that they usually
contain only zero or one galaxy, their counts-in-cells will have a
nearly Poisson distribution for which $b \simeq 0$ and
$\langle(\Delta N)^2\rangle \simeq \bar N$. Greater values of $b$
occur for larger departures from Poisson statistics, as equations
(\ref{1}) and (\ref{4}) indicate. Another way of visualizing this is
by relating $b$ to the two-galaxy correlation function:
\begin{equation}
b=-\frac{W}{2K}=\frac{2\pi Gm^{2} \bar n}{3T} \int _{V}\xi(\bar
n,T,r)\frac{1}{r}r^{2}dr \label{5}
\end{equation}
where $K$ is the kinetic energy and $T$ is the temperature given by
the peculiar velocity dispersion relative to the average Hubble
expansion. (For a detailed review see \citealt{sas00-b}.) For larger
volumes, there is a greater contribution of the two-galaxy
correlation function $\xi(r)$ to the integral of the gravitational
correlation energy $W$, and $b$ increases. On very large scales, the
correlations become  small and contribute no further to the integral
in equation (\ref{5}).  Thus the value of $b$ reaches its asymptotic
limit on these large scales, provided they are initially
uncorrelated. At low redshifts, $b(r)$ in the 2MASS catalog
can be used with equation (\ref{5}) to determine the two-galaxy
correlation function. Comparison of the result with the standard
direct determination shows excellent agreement \citep{siv05}.

The cell size $d (Mpc)$ is related to its angular size $\theta
$(radians) at a redshift $z$ by \citep {col02}
\begin{equation}
\theta =\frac{H_0 q^{2}_0}{c}~  \frac{(1+z)^{2}}{\left[q_0 z+(q_0
-1)((1+2q_0 z)^{0.5}-1) \right]} ~ d\label{6}
\end{equation}
where we take the Hubble constant $H_0 = 70$ $ km$ $ s^{-1}$ $
Mpc^{-1}$ and $q_0 = 0.5$.

\begin{figure}
\includegraphics[angle=0,clip=,width=3.20in]{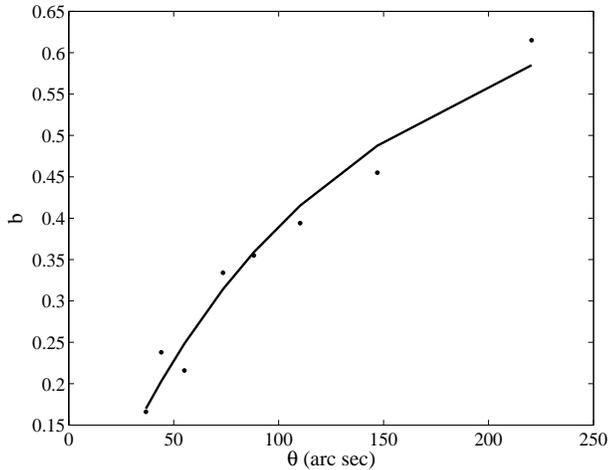} \caption{The points show the
$b$ values for different cell sizes in North with $0.47 < z < 0.8$.
The curve is the predicted $b(\theta)$ fitted on these points
according to
$b(\theta)=1-\left(1+\bar{N}s_0^{\gamma-1}C_{\gamma}\theta^{1-\gamma}\right)^{-1/2}$\citep{lah92}
with $s_0=0.0017^{\circ}$ and $\gamma=1.68$. \label{fig6}}
\end{figure}

Figure (\ref {fig5}) shows $b(\theta)$ for cells of different
angular sizes (in arc seconds) for the two redshift samples in the
North and South GOODS catalogs. Table (1) gives more detailed
information on these results. This information is useful for
understanding the physical scales of the cells and the numbers of
galaxies and cells being analyzed.  It also gives the numbers for
figure (\ref {fig4}) which are used to calculate the two-galaxy
correlation function (in figure (\ref{fig6})) from equation
(\ref{5}). Notice that for the same angular size, the physical
sizes of  cells in the South is slightly greater than in the North,
because of the slightly greater average redshift of the Southern
samples.  We have also examined the effect of the strong redshift
peak in the South at $z  = 0.74$ by reducing the upper redshift to
$z = 0.73$  which eliminates the peak in that nearer sample.  The
result is to decrease the values of $b$ for the same size cells by
between $0.01$ for the smallest cells and $0.1$ for cells of about
$220$ arc sec in length, usually an effect of about $15 - 20$\%.

Figure (\ref{fig6}) compares $b(\theta)$ with values based
on the two-galaxy angular correlation function, $W(s)$.
\citealt{lah92} derived $b(\theta)$ for square cells of size
$\omega=\theta \times \theta $ deg$^2$
\begin{equation}
b(\theta)=1-\left(1+\bar{N}s_0^{\gamma-1}C_{\gamma}\theta^{1-\gamma}\right)^{-1/2}\label{7}
\end{equation}
where $W(s)$ $=$ $(s/s_{0})^{1-\gamma}$, $\bar N$ is the expected
number of galaxies in a cell, and $C_{\gamma}$ is a coefficient to
be evaluated numerically. \citealt{lah92} calculated $C_\gamma$'s
corresponding to four different values of $\gamma=$ 1, 1.68, 1.8,
and 2 to be 1, 1.87, 2.25, and 2.97 respectively. We used the lower
redshift sample in the North (which seems  
the most homogeneous) and then fixed $C_\gamma$ to each of these
four values to fit equation (\ref{7}) to the data and
find $\gamma$ and $s_0$. As table (\ref{tab2}) shows by changing
$C_\gamma$ to its four numerically calculated values the fitted
values of $\gamma$ do not change but $s_0$ decreases.

To determine whether cells near the edges of each region could have
significant effects on the results, we omitted them and found that
they did not change the trend of the values of $b$ in the two
redshift intervals.

We also tested the effects of different cross-correlation radii
among the different catalogs which constitute the GOODS North and
South catalogs.  Increasing this cross-correlation radius from $0.5$
to $0.75$, $1.0$, and $1.5$ arcseconds gave $3009$, $3005$, $3003$
and $2994$ independent objects respectively.  So cross-correlation
radii in this range have negligible effects.

In the North Catalog, the majority of the data ($1536$ secure
redshifts) come from the TKRS redshift survey, and the other $434$
are from other redshift surveys.  To examine possible effects of
differences among these catalogs, we repeated all the analysis using
only the TKRS survey.  This gave small changes in the values of $b$
but did not affect the trend of values of $b$ between the two
redshift intervals.

As the total number of cells, $n$, in Table (1) decreases below
about $150$ (conservatively) the $f_V(N)$ histograms become less
smooth and their corresponding values of $b$ become less reliable.
The only ways to explore these regimes more accurately are with
large homogeneous samples covering greater areas, and with a larger
number of such samples.

\begin{deluxetable}{lrr}
\tablenum{2} \tablewidth{0pt}
\tablecaption{Two-point correlation function} \label{tab2}
\tablehead{\colhead{$c_{\gamma}$} & \colhead{$\gamma$} &
\colhead{$s_{0}$(degree)} \nl } \startdata 1    &  1.68 & 0.0017 \nl
1.87 &  1.68 & 0.0007 \nl 2.25 &  1.68 & 0.0005 \nl 2.97 &  1.68 &
0.0003 \nl
\enddata
\end{deluxetable}
\section{Discussion}
Our most striking result is that at redshifts up to about $z = 1.5$
the form of the spatial galaxy distribution function is remarkably
similar to its form at the present time.  Both these forms were
predicted by the gravitational statistical mechanics and
thermodynamics of the cosmological many-body system.  This indicates
that although merging and dark matter can be important for the
evolution of individual galaxies, they do not dominate the forms of
the large scale galaxy distribution.  The reasons for this lack of
dominance may be able to place important constraints on merging and
dark matter.

To obtain these constraints, we need to determine how $b(r,z)$
depends on the scale $r$, and how it evolves with redshift.  The
theory predicts both these dependencies (e.g. \citealt{sas00-b};
\citealt{ahm02}). In  particular $b$ should decrease with increasing
redshift.  Figure (\ref {fig5}) and Table (1) show that this
decrease holds in the North GOODS catalog, but not in the South. The
simplest explanation for this difference may be that the South
region is a compilation of four distinct catalogs, none of which
dominates. This may encourage more inhomogeneity than in the North
region which is dominated by one catalog.  Indeed
\citet{van05,van06} and \citet{rav07} have clearly noted the
inhomogeneity of structure in the South region. This may be a region
of excess clustering which would produce an unusually large apparent
local value of $b$.  (We recall that the usual value of $b$
represents an ensemble average over many regions which cover a
representative range of clustering.)  On the other hand, the
decrease of $b$ with $z$ in the North region behaves qualitatively
as expected.  Whether it agrees quantitatively with the
gravitational many-body theory is being investigated elsewhere
(\citet{yan08}).

When $b(r)$ can be determined more accurately, using a larger number
of larger homogeneous catalogs, it will also provide information on
the evolution of the two-galaxy correlation function.  (See
\citealt{siv05} for an example of the technique to accomplish this.)
This, in turn, can then be related to many computer simulations of
galaxy clustering which measure  the two-galaxy correlations under a
variety of conditions.
\acknowledgements Bill Saslaw thanks Naresh Dadhich, Ajit
Kembhavi and their colleagues for hospitality at the Inter
University Center for Astronomy and Astrophysics, as well as the
Cambridge Society of Bombay in Mumbai for its invitation to India.
Hadi Rahmani and Saeed Tavasoli thank Ajit Kembhavi for his
invitation to IUCAA and for useful discussions.



\end{document}